\documentstyle[11pt,aaspp4]{article}
\input{psfig}

\slugcomment{Accepted by ApJ Letters}
\lefthead{Bedding and Zijlstra}
\righthead{P-L relations for semiregulars and Miras}

\begin{document}

\title{Hipparcos period-luminosity relations for Miras and semiregular
variables}

\author{Timothy~R.~Bedding}
\affil{School of Physics, University of Sydney 2006, Australia}

\author{Albert~A.~Zijlstra}
\affil{UMIST, Department of Physics, P.O. Box 88, Manchester M60 1QD, UK}

\begin{abstract}
We present period-luminosity diagrams for nearby Miras and semiregulars,
selecting stars with parallaxes better than 20 per cent and well-determined
periods.  Using $K$-band magnitudes, we find two well-defined P-L
sequences, one corresponding to the standard Mira P-L relation and the
second shifted to shorter periods by a factor of about 1.9. The second
sequence only contains semiregular variables, while the Mira sequence
contains both Miras and semiregulars.  Several semiregular stars show
double periods in agreement with both relations.  The Whitelock
evolutionary track is shown to fit the data, indicating that the
semiregulars are Mira progenitors.  The transition between the two
sequences may correspond to a change in pulsation mode or to a change in
the stellar structure.  Large amplitude pulsations leading to classical
Mira classification occur mainly near the tip of the local AGB luminosity
function.

\end{abstract}

\keywords{stars: AGB and post-AGB}

\section{Introduction}

Long-period variables (LPVs) include Miras and semiregulars (SRs).  Mira
variables are located at the tip of the Asymptotic Giant Branch (AGB) where
they experience thermal pulses (\cite{I+R83}).  Semiregulars usually have
smaller amplitudes, shorter periods and more irregular pulsations than
Miras and they often show evidence for multiple periods.  The evolutionary
relation between the two groups is not clear but SRs are often considered
to be Mira progenitors, in which case they would be located lower on the
AGB\@.

Kerschbaum \&\ Hron (1992, 1994) have argued that some SRs have physical
characteristics identical to Miras, and are only excluded from the Mira
class because of the restrictive classical definition.  A good example is
the SRb star R~Dor, which Bedding et al.\ (1998) showed to be alternating
between periods of Mira and SR behaviour.  Such stars could be in a
transition phase, but it is also possible that Miras and SRs
co-exist during the same evolutionary phase, rather than forming a sequence
in evolution.

Miras in the LMC are known to follow a well-defined period--luminosity
(P-L) relation for both $M_{\rm bol}$ and $M_K$ (\cite{Fea96}).  This
implies that they all pulsate in the same mode, presumably radial, but
there is still controversy over whether this is the fundamental or first
overtone.  Because of their shorter periods, it is sometimes assumed that
SRs pulsate in higher overtones than Miras.  Support for this was found by
Wood \&\ Sebo\ (1996), who showed that LPVs in the Large Magellanic Cloud
(LMC) follow two sequences in a $K$-band P-L diagram.  One is the
well-known Mira sequence and the other (a sample of 8 stars) corresponds to
periods about two times shorter.  Sequences with shorter periods are seen
in MACHO observations of the LMC and the bulge (Minniti et al., 1998).  The
identification of semiregulars as higher overtone pulsators could mean that
Miras and SRs are in a very closely related evolutionary phase.

Results from Hipparcos for 16 of the nearest Mira variables were presented
by van Leeuwen et al.\ (1997) who showed that, with two exceptions, the
parallaxes are consistent with the same P-L relation as the LMC Miras.
They inferred that this P-L relation corresponds to first-overtone
pulsation, with the two stars that fall below the sequence ($\chi$~Cyg and
R~Cas) being fundamental-mode pulsators (see also Barth\`es\ 1998).  No
stars were found that might correspond to the LMC secondary sequence of
Wood \&\ Sebo\ (1996).

\section{Sample selection, periods and magnitudes}

We selected Miras and SRs from the Hipparcos Catalogue (\cite{ESA97}) using
the following criteria:
\begin{enumerate}

\item Parallax precision: $\sigma_\pi/\pi \le 0.2$.  For the Mira R~Leo, we
follow van Leeuwen et al.\ (1997) in adopting the weighted mean of the
Hipparcos parallax and a more precise ground-based measurement.

\item Coarse variability flag (catalogue field H6): $\ge2$.  This flag
indicates how much variability was measured from the Hipparcos photometry
(1: $<$0.06\,mag; 2: 0.06--0.6\,mag ; 3: $>$0.6\,mag).  This constraint
allows us to select true variables and to exclude M-dwarfs.

\item Spectral type (field H76): M* or S*.  We do not consider C-stars
here.
Hipparcos results for carbon LPVs have recently been discussed by Bergeat,
Knapik \& Rutily (1998).

\end{enumerate}

A fourth criterium was to select stars with well-defined periods longer
than 50 days.  The Hipparcos Catalogue gives periods in cases where a
solution was found in the photometric data, but the relatively short
timespan of observations means that periods for most SRs were not
identified.  The regular variations in these stars have small amplitudes
and are usually superimposed on larger and more irregular variations,
making a long time sequence necessary to determine the regular period.

We therefore used the General Catalogue of Variable Stars (GCVS; Kholopov
et al.\ 1988) to identify those stars with periods $\ge$50\,d.  For Miras
we adopted the GCVS periods, but for all other stars we insisted on
confirmation from an independent source.  In most cases, we analysed visual
observations supplied by the AFOEV (Association Francaise des Observateurs
d'Etoiles Variables) and the VSOLJ (Variable Stars Observers League in
Japan).  For six stars (RX Boo, T Cet, RS Cnc, AF Cyg, TX Dra and g Her) we
used the periods found by Mattei et al.\ (1997) from AAVSO data, while for
three others we adopted periods in the literature that we considered
reliable: R Dor (\cite{BZJ98}), RX Lep (\cite{CDS95}) and SW Vir (Armour,
Henry \& Baliunas 1990).  For many stars we could not locate sufficient
observations to confirm the GCVS period, while for others the observations
exist but do not reveal definite periodicity.  These stars were excluded
from the sample.

The final sample (Table~\ref{tab.hipp}) contains 6 Miras and 18
semiregulars, seven of which have two periods.  Note that two stars have
been classified in the GCVS as SRc (Column 2), but this classification is
reserved for supergiants and these stars should be re-classified SRb.  One
star (W~Hya) is classified SRa, reflecting its very regular pulsations.

We have taken $K$ magnitudes for the Miras from van Leeuwen et al.\ (1997).
For most SRs, we used values in Kerschbaum \&\ Hron (1994, reduced
to the Carter 1990 system by subtracting 0.031\,mag), the IRC Catalogue and
a few other references in the SIMBAD database.  For R~Dor we used the mean
$K$ magnitude from Bedding et al.\ (1997) and for W~Hya we used a mean
value from Whitelock (private communication).  For many SRs these
are single-epoch observations.  Stellar variability will introduce some
additional scatter into the P-L diagram, but the $K$-band amplitudes of SRs
are small.  Even for Miras, single-phase observations produce a scatter
about the mean $(K, \log P)$ relation in the LMC of only 0.26\,mag
(\cite{H+W90}) and we expect the spread for SRs to be less than 0.1\,mag.
For SRs for which we have multiple $K$-band measurements, these
generally agree to within 0.1\,mag.

\subsection{Bias}

Our sample is selected on the basis of $\sigma_\pi/\pi$ so it will be
affected by Lutz-Kelker bias (\cite{L+K75,Koe92}, Oudmaijer, Groenewegen \&
Schrijver): on average, distances will tend to be underestimated.  This is
partially cancelled by the decrease of precision at fainter magnitudes,
which causes us to select fewer stars at large distances.  We can use
calculations by Koen\ (1992) to estimate the net bias in absolute
magnitude, where we select the $p=2$ case so as to include both effects.
The most probable value for the bias is $-$0.05\,mag for stars with
$\sigma_\pi/\pi = 15\%$ and $-$0.1\,mag for $\sigma_\pi/\pi = 20\%$.  The
mean values for the expected bias are $-$0.2\,mag and $-$0.4\,mag,
respectively, although individual confidence limits are much poorer.  The
above values give estimates of the most likely and average amounts by which
stars have been shifted down in the P-L relations (being slightly more
luminous than indicated by their parallaxes).

In summary, even for the least accurate parallax measurements in our
sample, the sample bias in $M_K$ is most likely around $-$0.1\,mag, which
is comparable to the scatter in the photometry.  For some individual stars,
the effect will be larger but is still not likely to change the conclusions
of this paper, particularly the existence of two separated P-L sequences.

\section{Discussion}

\subsection{Mira and semiregular relations}

Figure~\ref{fig.pl-all} shows the $K$-band P-L diagram for stars in our
sample (stars with two periods are plotted twice).  The solid diagonal
line shows the LMC P-L relation of Feast\ (1996):
\begin{equation}
  M_K = -3.47 \log P + 0.91,
\end{equation}
where $P$ is the period in days and we adopt a distance modulus for the LMC
of 18.56.  The crosses show the LMC stars on which the relation is based
and the asterisks show the LMC cluster LPVs of Wood \&\ Sebo\ (1996). The
two sequences defined by the LMC stars agree well with our sample of
galactic LPVs.

Figure~\ref{fig.hist} shows the histogram of vertical distance from the P-L
relation for our sample.  There is a peak at zero, containing stars in
agreement with the relation, and another about 0.9\,mag above it
(corresponding to a leftwards shift in period by factor of 1.8).  This
confirms the existence of the secondary relation reported by Wood \&\ Sebo\
(1996).  In addition, several stars clearly fall below the relation by
about 1\,mag.  The six Miras in our sample all fall either on the P-L
relation or below (for convenience, in Fig.~\ref{fig.pl-names} we reproduce
the P-L diagram of our sample with each point labelled).  In contrast, all
stars lying above the relation are classified as SR\@.

Of the seven SRs which have two periods, in five cases the longer period
falls on the standard Mira relation and the shorter falls on or near the SR
relation.  (TX~Dra and W~Cyg do not give such good agreement, but their
long periods only differ from Mira relation by less than 2$\sigma$.
Furthermore, their period ratios agree well with those of the other stars.)
The observed period ratios in these stars (1.76--1.90) should provide
useful constraints for pulsation models.

\subsection{The Whitelock evolutionary track}

The Mira P-L relation can not be an evolutionary sequence (e.g., Whitelock,
Feast \& Catchpole 1991).  Instead, Whitelock (1986) has shown that Miras
and SRs within a globular cluster define a sequence in the P-L diagram
which is shallower than the Mira P-L relation.  Within each cluster, Miras
are located at the intersection of this sequence with the P-L relation.
This sequence, the Whitelock track, is therefore the most probable
evolutionary track.  The slope of the Whitelock track agrees with
evolutionary calculations of Vassiliadis \&\ Wood (1993).

The Whitelock track has only been defined for $M_{\rm bol}$. To convert it
to $M_K$, we used data from Whitelock (1986) for 47 Tuc and NGC 5927, two
clusters with similar high metallicity and Mira populations with the same
period distribution.  For these stars, we find that the Whitelock track is
fitted by:
\begin{equation}
  M_K = -(1.67\pm0.12)\log P -(3.05\pm0.25), 
\end{equation}
where the zero point depends on the distance scale used.  The slope of this
Whitelock track varies little for stars with initial masses up to
2.5\,$M_{\sun}$ (\cite{V+W93}), while the end point (lying on the Mira P-L
relation) shifts to brighter magnitudes for younger populations.

The dotted line in Fig.~\ref{fig.pl-all} shows the Whitelock track shifted
up by 0.8\,mag so as to pass through the SRs.  The track fits reasonably
well and connects the highest density of points on the P-L relation with
the same for the SR sequence, confirming its importance as an evolutionary
track.  The SRs in our sample can thus be considered to be progenitors of
Mira having periods above 300 days.  This agrees with kinematical studies
of SRs (Feast, Woolley \& Yilmaz 1972), which show a relatively low
velocity dispersion, independent of period and similar to the velocity
dispersion of long-period Miras.  Shorter-period Miras show higher velocity
dispersions, indicating an older population.  Hipparcos proper motions for
our sample also agree with these findings.

Although the shifted Whitelock track fits the SR data quite well, it does
not explain the clustering into two separate sequences.  The clustering
shows that the evolution along the Whitelock track is not continuous:
instead, stars spend more time near the location of the sequences.  This,
together with the double-mode stars that may fit both sequences, can be
interpreted as a difference in pulsation mode between the two sequences.
In this interpretation, the luminosity of the star increases during the SR
phase, followed by a sudden increase in period due to a mode change, after
which evolution continues on the Mira relation.  However, this
interpretation is by no means proven.  An alternative possibility is that
the period changes due to an adjustment in the stellar structure.  In this
case, the fact that the shorter periods of double-mode LPVs fit the SR
sequence would be coincidental.

\section{Conclusions}

We have identified period--luminosity sequences for Miras and semiregulars
by selecting stars with precise Hipparcos parallaxes.  Our main conclusions
are as follows:
\begin{enumerate}

\item A significant number of SRs fall on the standard Mira P-L
relation, implying that they pulsate in the same mode as most Miras
and are closely related in evolution.

\item The remaining SRs define a second P-L sequence with the same
slope but shifted to shorter periods by a factor of about 1.9.  Most
of the SRs that fall on the standard Mira P-L relation have a
secondary period which falls on or near the SR relation.  The period
ratios are in the range 1.76--1.9.

\item Large amplitude pulsations seen in classical Miras occur mainly near
the tip of the local AGB luminosity function.

\item The slope of the Whitelock evolutionary track agrees with the data,
implying that the SRs are progenitors of long-period Miras. The separation
into two sequences may be due to either a difference in pulsation mode, or
to an adjustment in the stellar structure.

\end{enumerate}

\section*{Acknowledgments}

We are grateful to the Hipparcos team and to the hundreds of amateur
observers whose measurements were critical to this paper.  Visual data for
many stars were obtained from the AFOEV database, operated at CDS, France
and the VSOLJ database in Japan.  We also made extensive use of the SIMBAD
and ADS data services.  We thank the referee, Mike Feast, for valuable
suggestions which led to significant improvements in the paper.  For
financial support, TRB is grateful to the Australian Research Council and
AAZ thanks the European Southern Observatory.

\clearpage

\centerline{\psfig{file=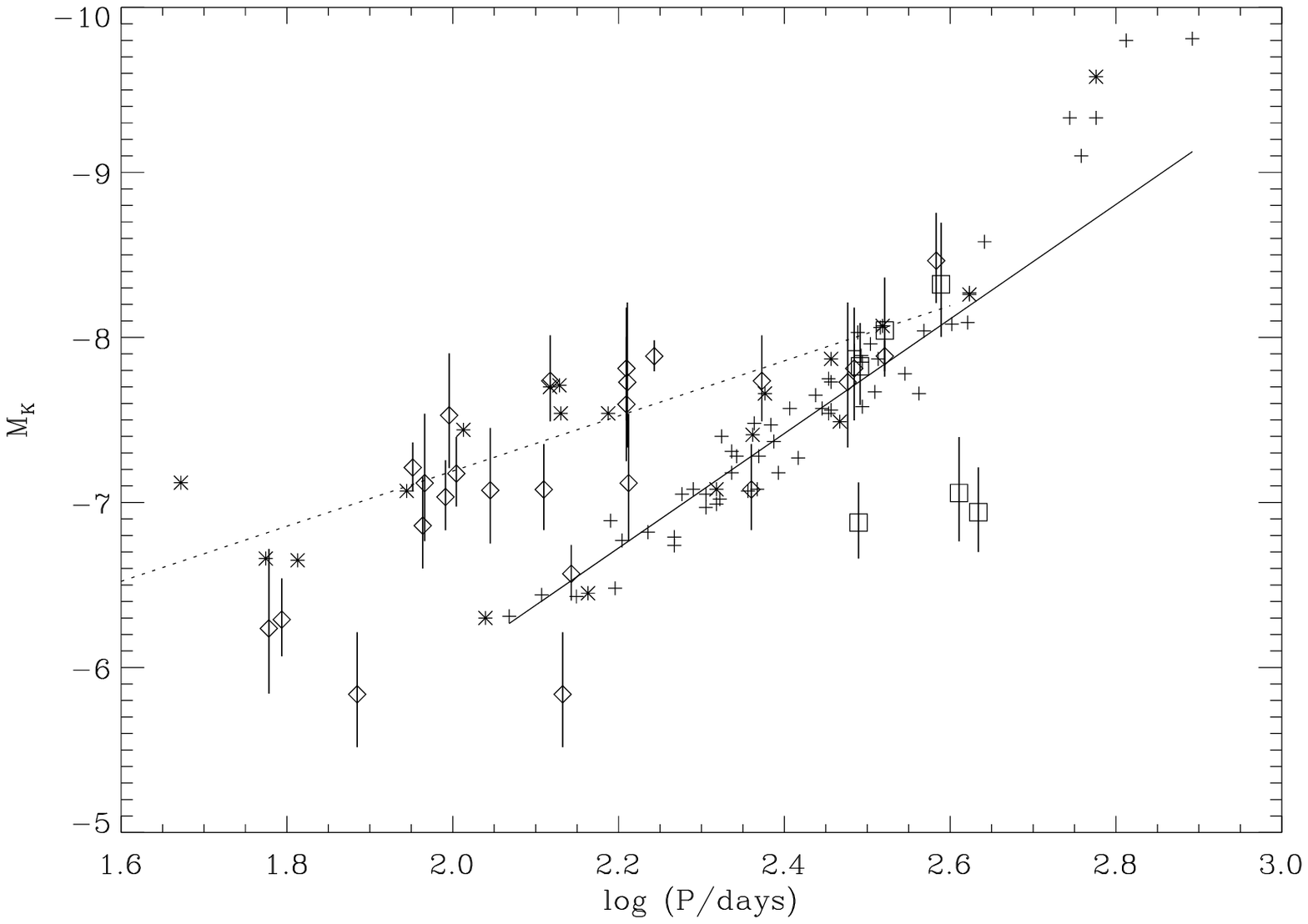}}

\figcaption[]{\label{fig.pl-all} The positions of LPVs in the $K$-band
period-luminosity diagram.  Points with error bars are the stars in
Table~\ref{tab.hipp} (squares are Miras and diamonds are semiregulars).
The solid line shows the LMC Mira relation from Feast\ (1996) and the
crosses show the data on which it is based (\cite{FGW89}).  The asterisks
show LMC cluster LPVs from Wood \&\ Sebo\ (1996).  The dotted line shows
the Whitelock track shifted upwards by 0.8\,mag (see text).  }

\centerline{\psfig{width=11cm,file=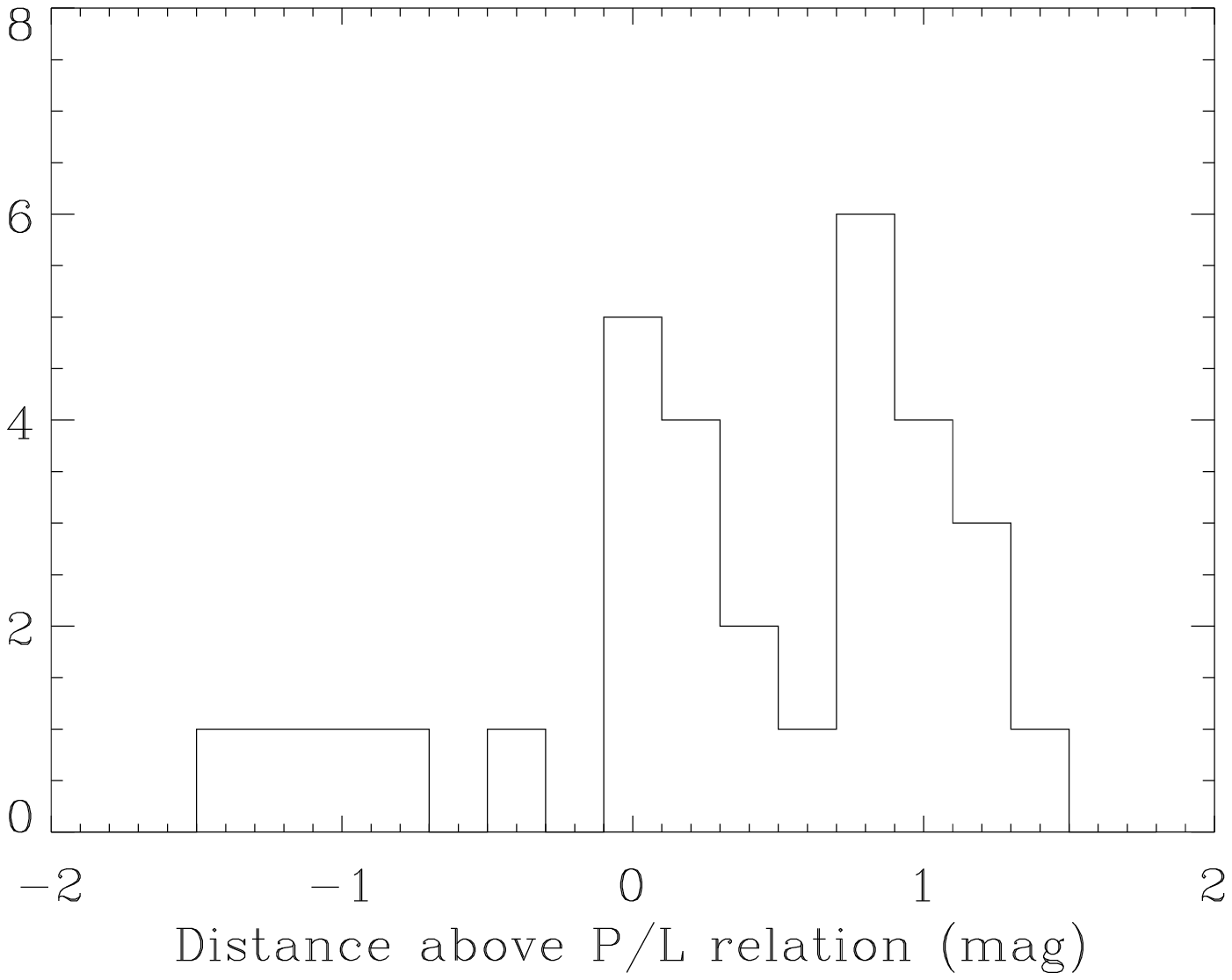}}

\figcaption[]{\label{fig.hist} Histogram of vertical distance from the
standard Mira P-L relation for the stars in our sample.}

\centerline{\psfig{width=15cm,file=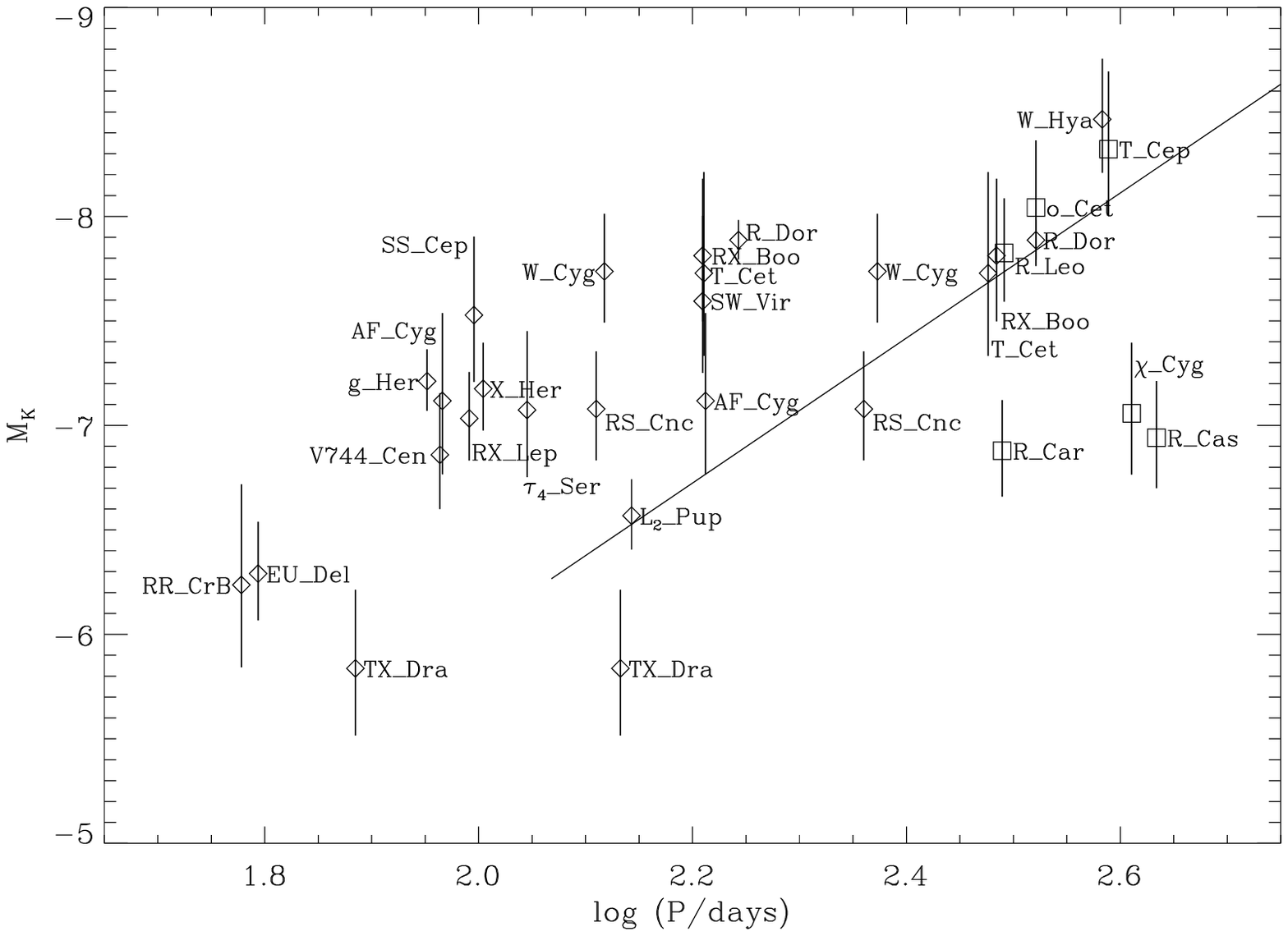}}

\figcaption[]{\label{fig.pl-names} The P-L diagram for our sample, with all
points labelled.  }

\clearpage

\begin{deluxetable}{rlrrrrr}
\footnotesize
\tablecaption{\label{tab.hipp} Sample of long-period variables}
\tablewidth{0pt}
\tablehead{
\colhead{Name} & \multicolumn{2}{c}{GCVS period} & \colhead{$\pi$} &
\colhead{$\sigma_\pi$} & \colhead{$K$} &
\colhead{Period(s)}\\
               &                  & \colhead{(days)} &
\colhead{(mas)} & \colhead{(mas)} & \colhead{(mag)} &
\colhead{(days)}}
\startdata
RX       Boo  & SRb  & 340\rlap{:}&  6.42 & 1.00 & $-$1.85 & 162, ~305 \\
R        Car  & M    & 308.7      &  7.84 & 0.83 & $-$1.35 & 309 \\
R        Cas  & M    & 430.5      &  9.37 & 1.10 & $-$1.80 & 430 \\
V744     Cen  & SRb  & ~90\rlap{:}&  6.00 & 0.76 & $-$0.75 & ~92 \\
T        Cep  & M    & 388.1      &  4.76 & 0.75 & $-$1.71 & 388 \\
SS       Cep  & SRb  & ~90        &  4.04 & 0.64 & $-$0.56 & ~99 \\
T        Cet  & SRc  & 158.9      &  4.21 & 0.84 & $-$0.85 & 162, ~299 \\
$o$      Cet  & M    & 332.0      &  7.79 & 1.07 & $-$2.50 & 332 \\
RS       Cnc  & SRc: & 120\rlap{:}&  8.21 & 0.98 & $-$1.65 & 129, ~229 \\
RR       CrB  & SRb  &  60.8      &  3.67 & 0.73 &    0.94 & ~60.0 \\
W        Cyg  & SRb  & 131.1      &  5.28 & 0.63 & $-$1.35 & 131, ~236 \\
AF       Cyg  & SRb  &  92.5      &  3.30 & 0.58 &    0.29 & ~92.5, ~163 \\
$\chi$   Cyg  & M    & 408.1      &  9.43 & 1.36 & $-$1.93 & 408   \\
EU       Del  & SRb  &  59.7      &  9.16 & 0.99 & $-$1.10 & ~62.2 \\
R        Dor  & SRb  & 338\rlap{:}& 16.02 & 0.69 & $-$3.91 & 175, ~332 \\
TX       Dra  & SRb  &  78\rlap{:}&  3.52 & 0.56 &    1.43 & ~77, ~136 \\
X        Her  & SRb  &  95.0      &  7.26 & 0.70 & $-$1.48 & 101   \\
g        Her  & SRb  &  89.2      &  9.03 & 0.61 & $-$1.99 & ~89.5 \\
W        Hya  & SRa  & 361        &  8.73 & 1.09 & $-$3.17 & 383   \\
R        Leo  & M    & 310.0      &  8.81 & 1.00 & $-$2.55 & 310   \\
RX       Lep  & SRb  &  60\rlap{:}&  7.30 & 0.71 & $-$1.35 & ~98   \\
L$_2$    Pup  & SRb  & 140.6      & 16.46 & 1.27 & $-$2.65 & 139   \\
$\tau^4$~Ser  & SRb  & 100\rlap{:}&  6.27 & 1.00 & $-$1.06 & 111   \\
SW       Vir  & SRb  & 150\rlap{:}&  7.00 & 1.20 & $-$1.82 & 162   \\
\enddata
\end{deluxetable}

\end{document}